\documentclass[a4paper]{appolb}
\usepackage{graphicx}
\usepackage{hyperref}
\usepackage{amsmath}
\usepackage{bm}

\usepackage{cite}
\usepackage{booktabs}
\newcommand{\trento}{T\raisebox{-0.3ex}{R}ENTo}

\begin{document}
\title{Comprehensive Study of \\Multi-scale Jet-medium Interaction\thanks{Presented at the 29th International Conference on Ultrarelativsitic Nucleus-Nucleus Collisions}
}
\author{Yasuki Tachibana${}^{1}$ on behalf of the JETSCAPE Collaboration
\address{${}^{1}$Akita International University
}
}
\maketitle
% NOTE: There was a citation to the qhat hadron RAA paper, but the journal consistently
%       asks to remove citations from the abstract. So I'll just remove it here.
\begin{abstract}
We explore jet-medium interactions at various scales in high-energy heavy-ion collisions using the \texttt{JETSCAPE} framework. 
The physics of the multi-stage modeling and the coherence effect at high virtuality is discussed through the results of multiple jet and high-$p_{\mathrm{T}}$ particle observables, compared with experimental data. 
Furthermore, we investigate the jet-medium interaction involved in the hadronization process.
\end{abstract}

\hypertarget{introduction}{%
\section{Introduction}\label{introduction}}
In jet-shower evolution, the virtuality and energy of each jet parton vary considerably. 
Thus, in high-energy heavy-ion collisions, jets can be used as dynamical probes to investigate the jet-medium interaction at various scales. 
\texttt{JETSCAPE} ~\cite{JETSCAPE:2017eso, JETSCAPE:2019udz, JETSCAPE:2020shq, JETSCAPE:2020mzn, JETSCAPE:2021ehl, JETSCAPE:2022jer, JETSCAPE:2022hcb}
is a publicly available software framework for Monte Carlo event generators that enables simulations describing physics at varying scales involved in in-medium jet evolution. 
The \texttt{JETSCAPE} framework incorporates multiple models, each effective at an individual scale range, and switches between them at appropriate scales while mediating their communication.

As a new feature, the jet quenching strength $\hat{q}$ with an explicit virtuality dependence due to the resolution scale evolution of jets~\cite{Kumar:2019uvu} is now supported by \texttt{JETSCAPE}. 
In these proceedings, we demonstrate that this further extension is crucial for a simultaneous description of the nuclear modification factor for inclusive jets and leading hadrons. 
Observables for jet substructures and heavy-flavor jets are also explored for more detailed discussions of the virtuality dependence in the jet-medium interaction. 
Furthermore, we present the effect of jet-medium interaction in the hadronization process through a systematic study using the \texttt{Hybrid Hadronization} module in the \texttt{JETSCAPE} package.

\hypertarget{Multi-stage jet evolution in \texttt{JETSCAPE}}{%
\section{Multi-stage jet evolution in JETSCAPE}\label{jet}} 
We present results from a multi-stage jet evolution model, the \texttt{MATTER} energy loss module handles partons with large virtualities, and the \texttt{LBT} energy loss module is applied to nearly on-shell partons within the \texttt{JETSCAPE} framework. 
In the \texttt{MATTER} phase, the coherence effects by the scale evolution of the medium parton distribution~\cite{Kumar:2019uvu} are implemented as the effective jet quenching strength $\hat{q} = \hat{q}_\mathrm{HTL}f(Q^2)$ with the virtuality dependent modulation factor parametrized as~\cite{JETSCAPE:2022jer}
\begin{align}
f(Q^2) & = \frac{1+10\ln^{2}(Q^2_\mathrm{sw}) + 100\ln^{4}(Q^2_\mathrm{sw})}{1+10\ln^{2}(Q^2) + 100\ln^{4}(Q^2)},
\label{eq:qhatSuppressionFactor}
\end{align}
where $Q^2_\mathrm{sw}$ is the switching virtuality between \texttt{MATTER} and \texttt{LBT}. 
If one eliminates the virtuality dependence ($f(Q^2)=1$), the effective jet quenching strength is reduced to the jet quenching parameter for on-shell parton calculated by the hard-thermal-loop (HTL) effective theory~\cite{He:2015pra}. 
The space-time medium profile for the energy loss calculations is obtained through (2+1)-D freestreaming~\cite{Liu:2015nwa} and subsequent (2+1)-D viscous hydrodynamic evolution by \texttt{VISHNU} ~\cite{Shen:2014vra} with the initial condition by \texttt{\trento}\ \cite{Moreland:2014oya}. 
The values of free parameters are the same throughout all calculations. 

\begin{figure}[htb]
\begin{center}

\vspace{10pt}

\includegraphics[width=0.5\textwidth]{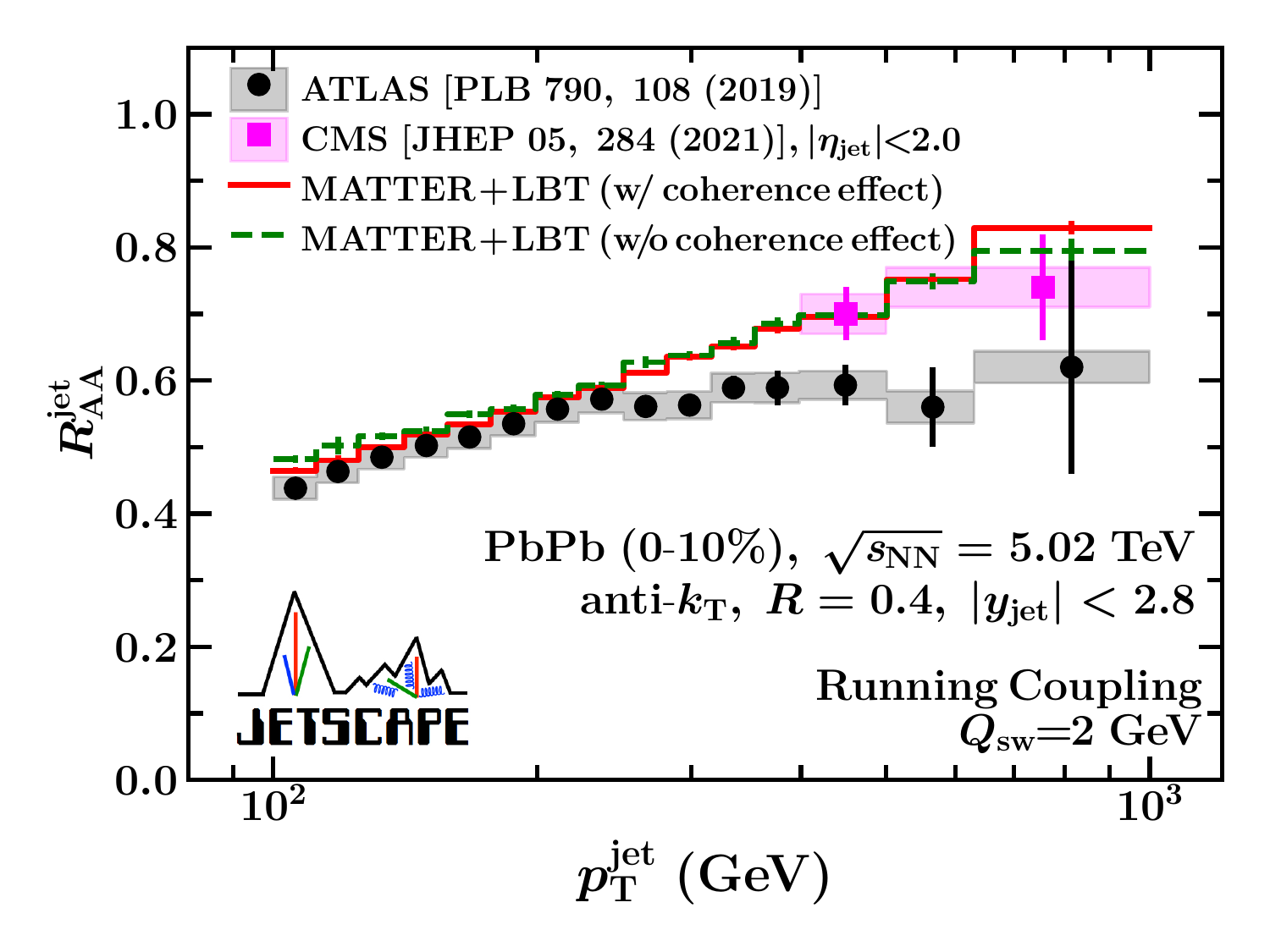}
\hspace{-1.25em}
\includegraphics[width=0.5\textwidth]{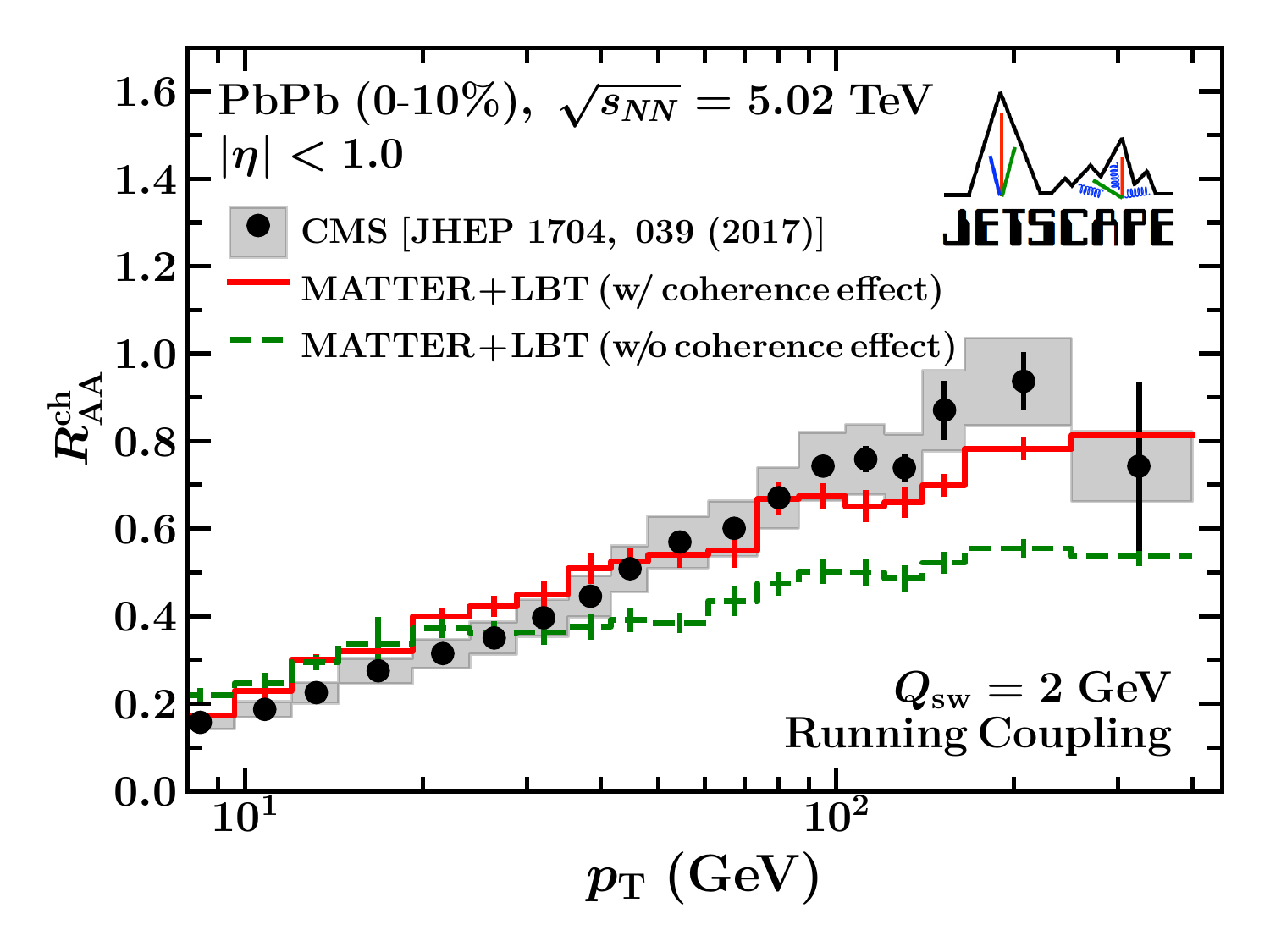}

\includegraphics[width=0.5\textwidth]{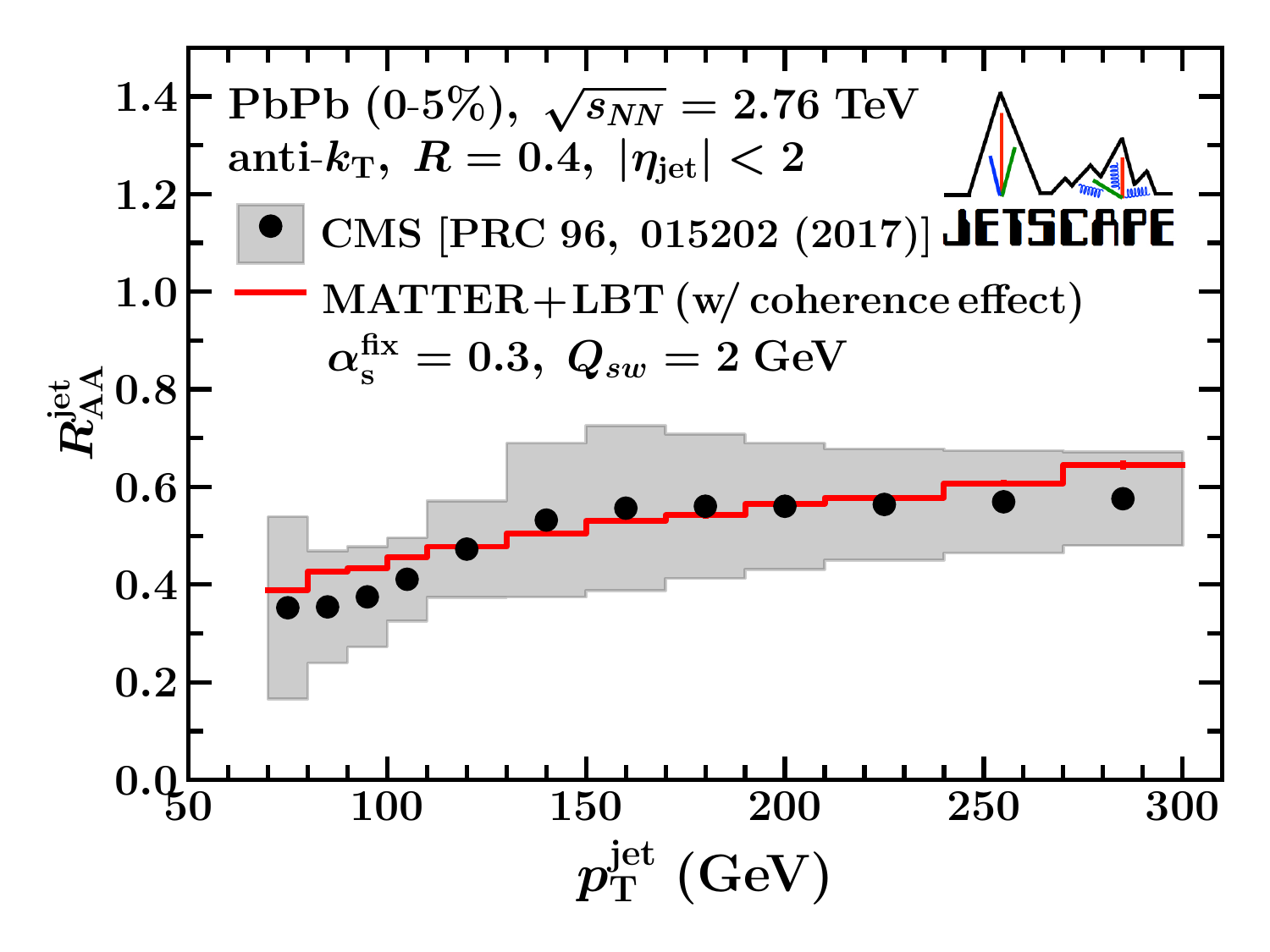}
\hspace{-1.25em}
\includegraphics[width=0.5\textwidth]{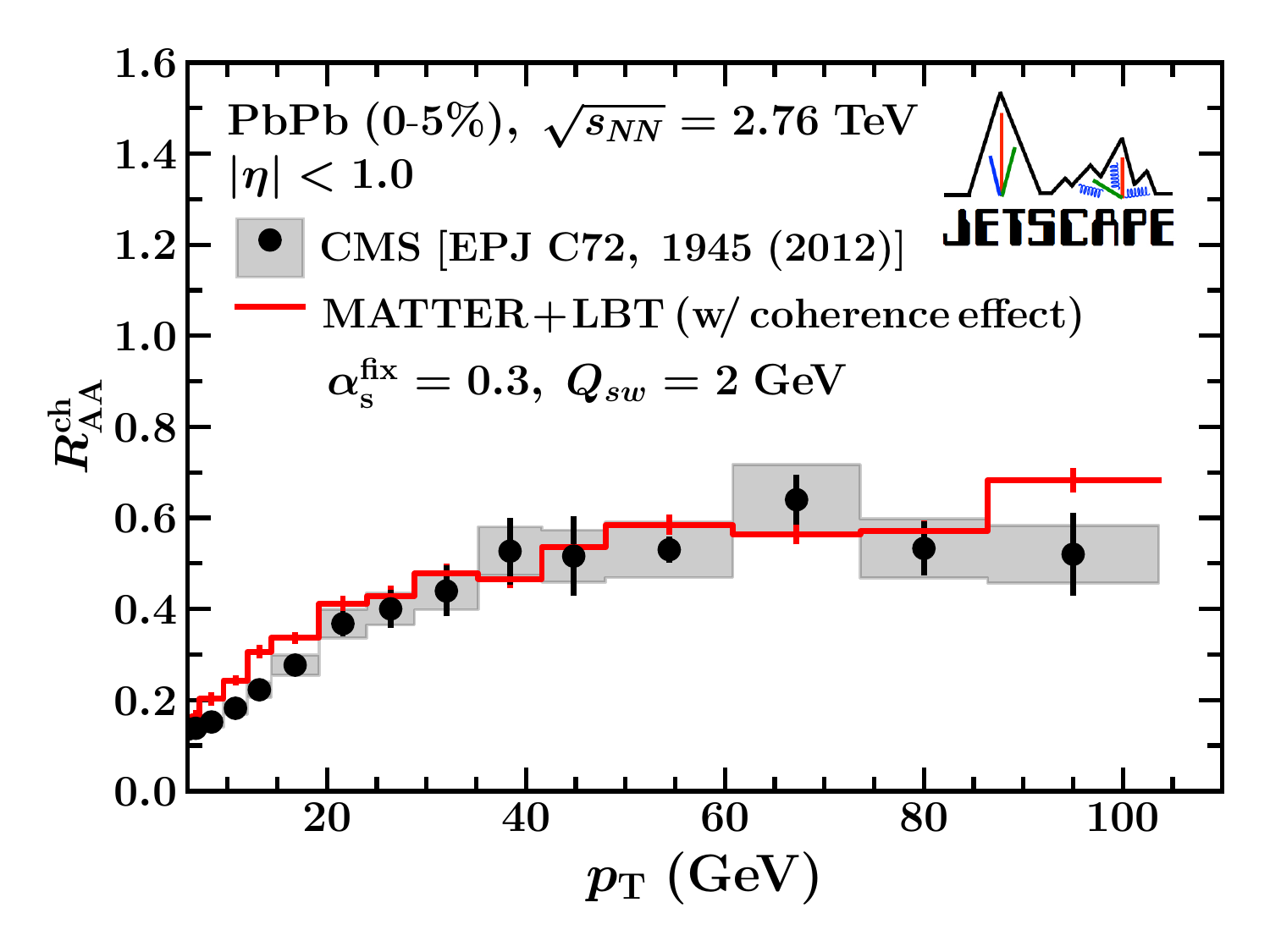}

\includegraphics[width=0.5\textwidth]{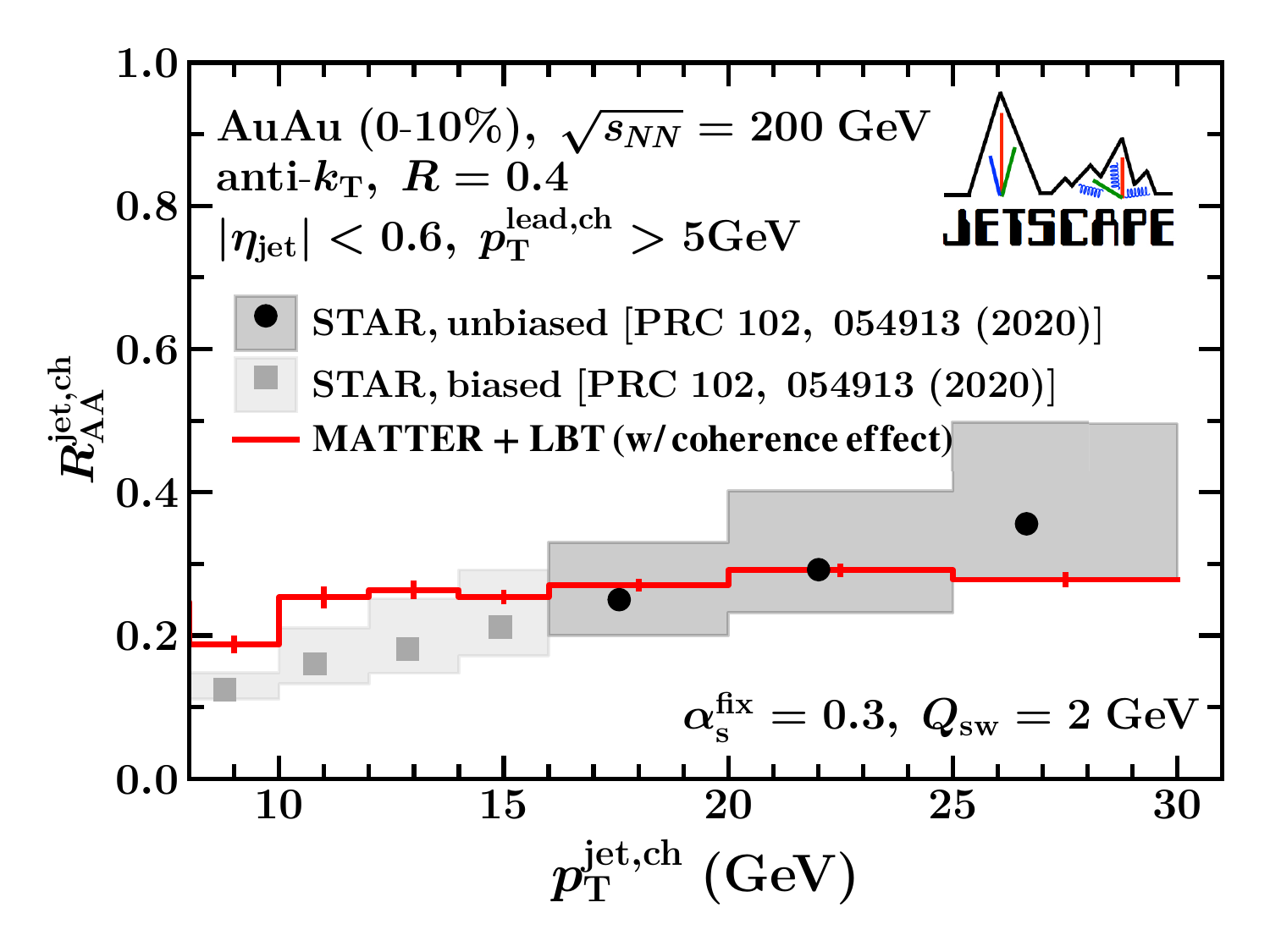}
\hspace{-1.25em}
\includegraphics[width=0.5\textwidth]{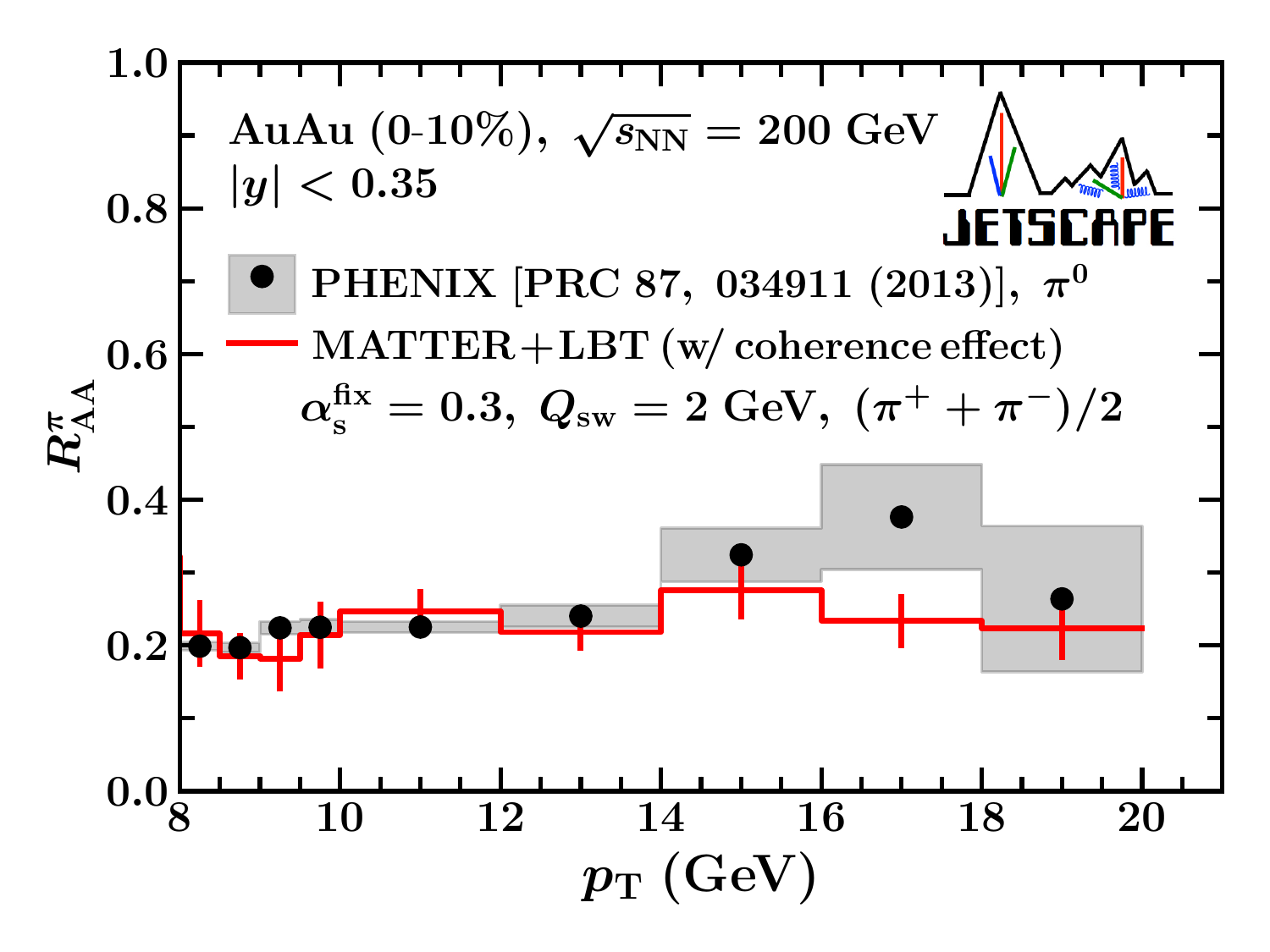}

\end{center}
%\vspace{-16pt}
\caption{Nuclear modification factors for reconstructed jet (left panels) and single particle (right panels). 
Top: Pb+Pb collisions at $5.02$~TeV. 
Middle: Pb+Pb collisions at $2.76$~TeV. 
Bottom: Au+Au collisions at $200$~GeV. 
Experimental data are taken from ATLAS~\cite{ATLAS:2018gwx}, CMS~\cite{CMS:2021vui,CMS:2016xef,CMS:2016uxf,CMS:2012aa}, STAR~\cite{STAR:2020xiv}, and PHENIX~\cite{PHENIX:2012jha}. 
}
\label{fig:raa}
\end{figure}

\begin{figure}[htb]
\begin{center}
\includegraphics[width=0.95\textwidth]{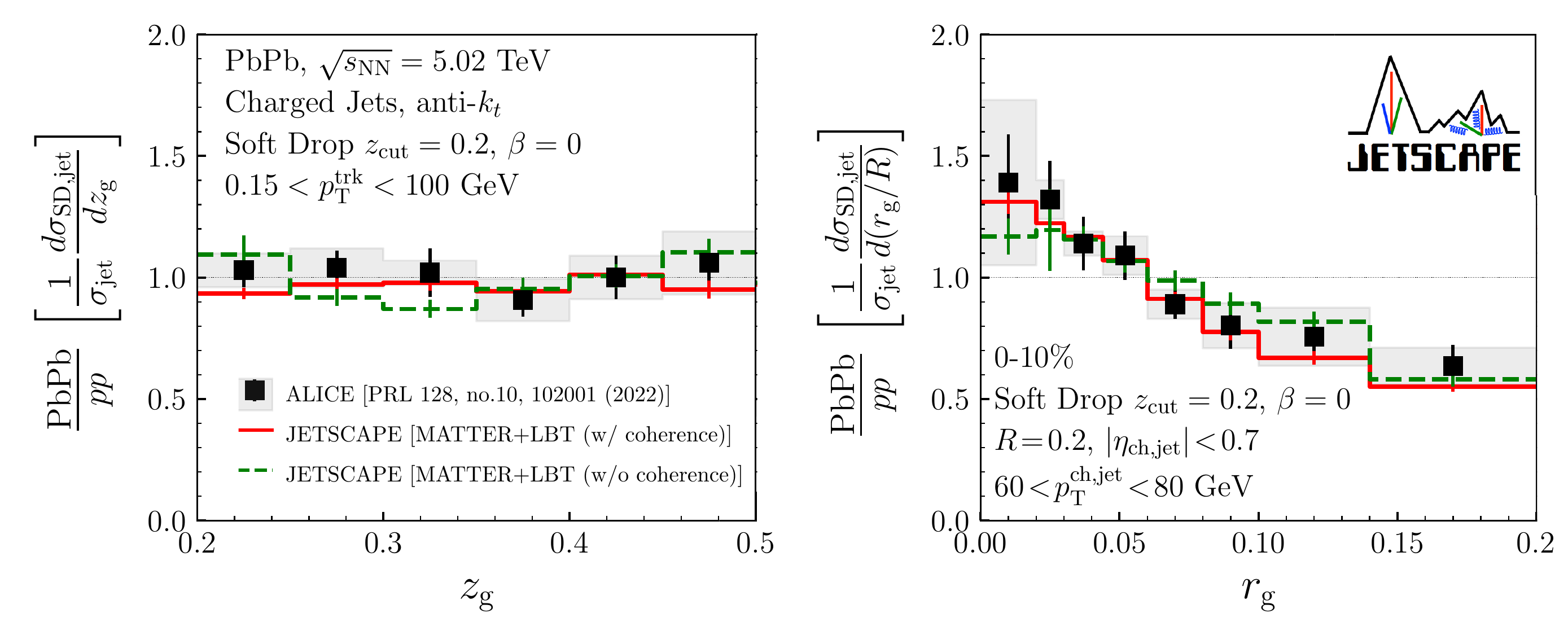}
\end{center}
\vspace{-12pt}
\caption{Modification of $z_g$ (left) and $r_g$ (right) distribution for charged jet in Pb+Pb collisions at $5.02$~TeV. Experimental data are taken from ALICE\cite{ALargeIonColliderExperiment:2021mqf}. 
}
\label{fig:groomed}
\end{figure}

Figure~\ref{fig:raa} shows the nuclear-modification factor for the reconstructed jet and single particle. 
Our full results with the coherence effects describe the experimental data from the top RHIC to the top LHC collision energies well.  
In particular, the coherence effects are significant in describing the high-$p_{\mathrm{T}}$ single particle at $5.02$~TeV. 
Besides, excellent descriptions of the groomed jet observables measured by ALICE are presented in Fig.~\ref{fig:groomed}. 

\begin{figure}[htb]
\begin{center}
\includegraphics[width=0.9\textwidth]{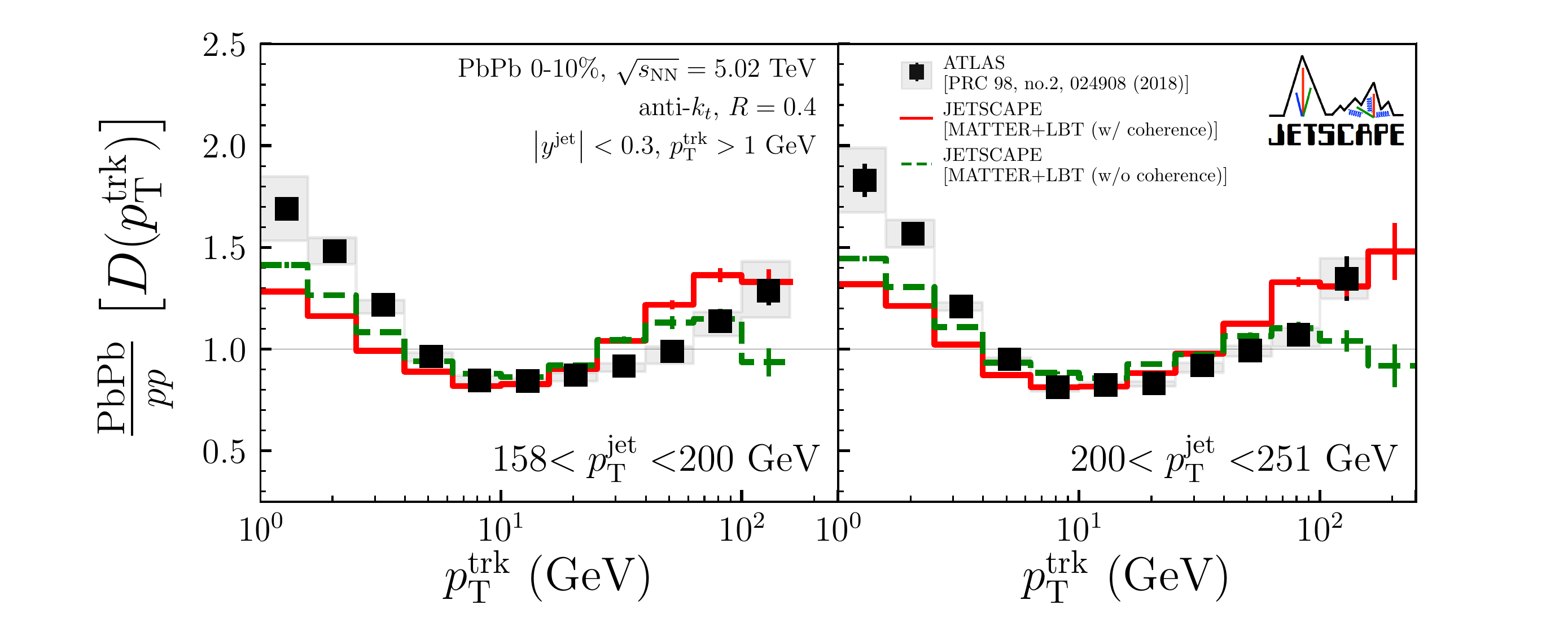}
\end{center}
\vspace{-12pt}
\caption{Modification of jet fragmentation function in Pb+Pb collisions at $5.02$~TeV. Experimental data are taken from ATLAS~\cite{ATLAS:2018bvp}.
}
\label{fig:ff}
\end{figure}

The coherence effects also manifest in the jet substructures, such as the jet-fragmentation function shown in Fig.~\ref{fig:ff}. 
Since the energy loss of partons in the jet core is suppressed by the coherence effect, 
the enhancement of high-$p_{\mathrm{T}}$ constituents can be seen. 

\begin{figure}[htb]
\begin{center}

\includegraphics[width=0.46\textwidth]{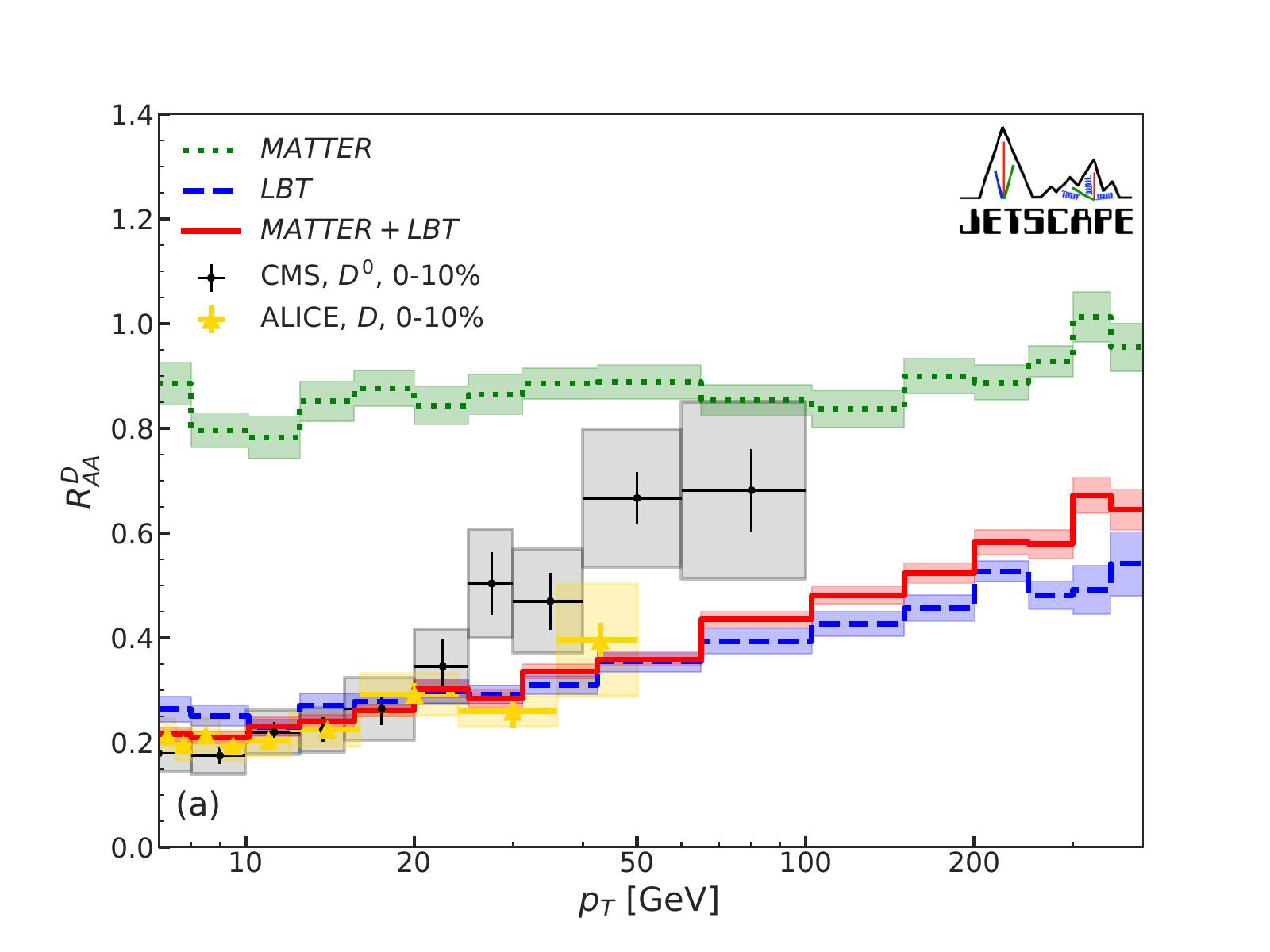}
\hspace{-2.em}
\includegraphics[width=0.46\textwidth]{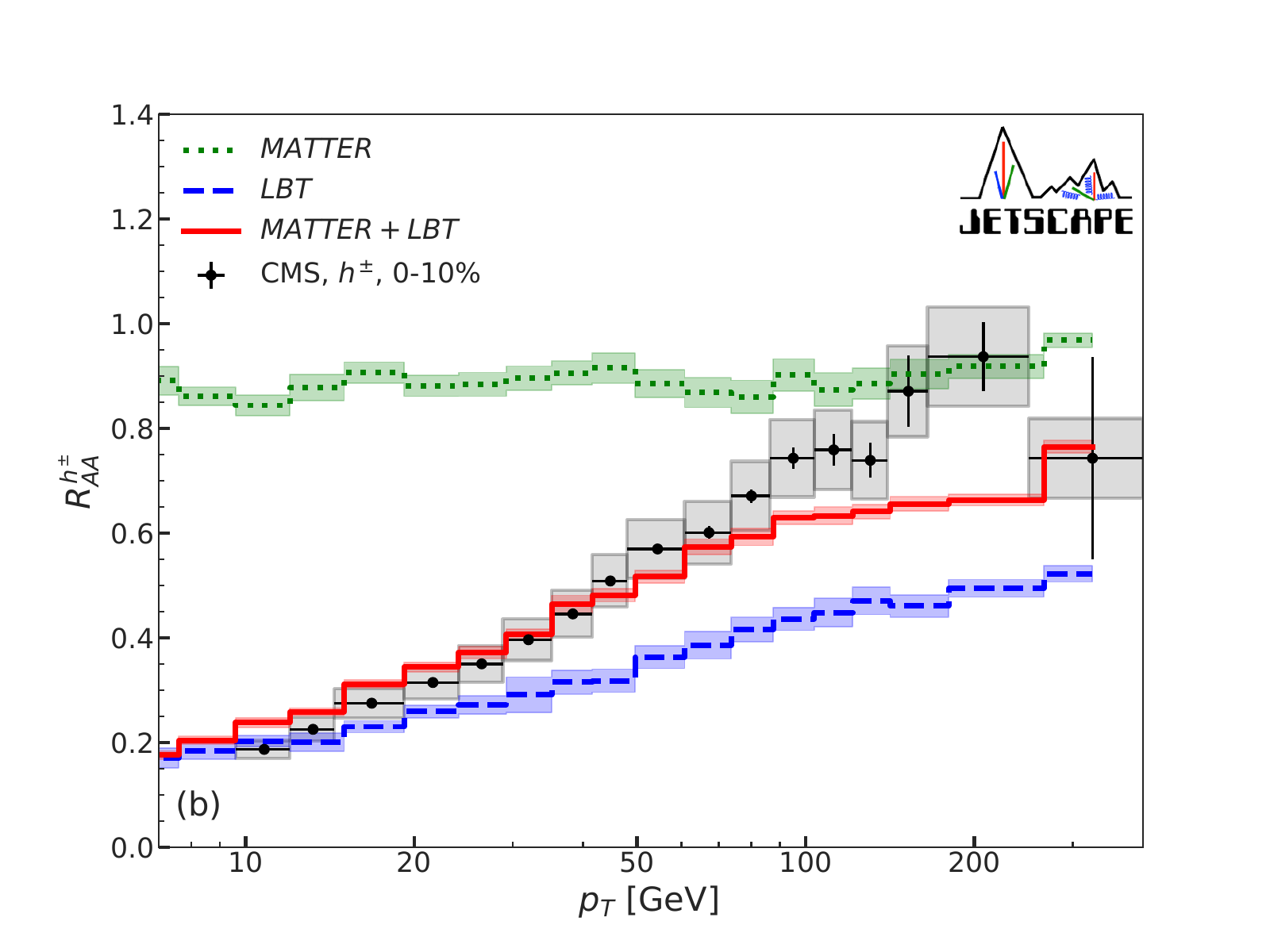}

\end{center}
\vspace{-12pt}
\caption{Nuclear modification factors for $D^0$ meson (left) and inclusive charged hadron (right). 
Experimental data are taken from CMS~\cite{CMS:2017qjw,CMS:2016xef}, and ALICE~\cite{ALICE:2021rxa}. 
}
\label{fig:charm}
\end{figure}

The contribution from each phase of the multi-stage description in charm and inclusive hadron energy loss is shown in Fig.~\ref{fig:charm}. 
It can be seen that the low virtuality phase dominates the charm quark energy loss compared to the light flavor partons. 
%%%%%%%%%%%%%%%%%%%%%%%%%%%%%%%%%%%%%%%%%%%%

%%%%%%%%%%%%%%%%%%%%%%%%%%%%%%%%%%%%%%%%%%%%
%%%%%%%%%%%%%%%%%%%%%%%%%%%%%%%%%%%%%%%%%%%%
%%%%%%%%%%%%%%%%%%%%%%%%%%%%%%%%%%%%%%%%%%%%
%%%%%%%%%%%%%%%%%%%%%%%%%%%%%%%%%%%%%%%%%%%%

\newpage

%%%%%%%%%%%%%%%%%%%%%%%%%%%%%%%%%%%%%%%%%%%%
\hypertarget{Jet-medium interaction in hadronization}{%
\section{Jet-medium interaction in hadronization}\label{hybrid}}
\texttt{Hybrid Hadronization}~\cite{Fries:2019vws} is a model which interpolates string fragmentation in a vacuum and quark recombination in a dense QGP medium. 
Its comprehensive description even includes recombinations of jet partons with thermal medium partons. 

\begin{figure}[htb]
\begin{center}
\includegraphics[width=1\textwidth]{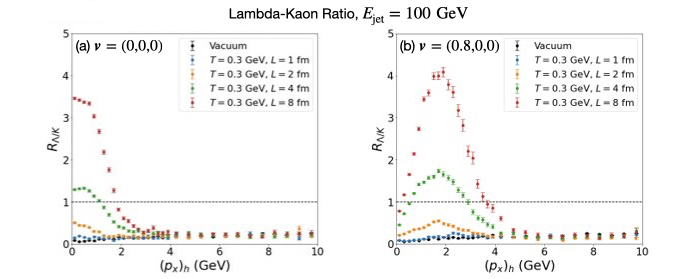}
\end{center}
\caption{Yield ratios of $\varLambda$ to $K$ in jets as a function of momentum in $x$ direction for the brick medium with flow velocity (a) $\bm{v}=(0,0,0)$, and (b) $\bm{v}=(0.8,0,0)$. 
Each color of the markers represents the path length of the jets in the medium $L$. }
\label{fig:hybrid}
\end{figure}

Figure~\ref{fig:hybrid} shows $\varLambda$-to-$K$ ratio of longitudinal momentum spectra in a jet. 
Here a jet shower with $E=100$~GeV traveling in the $x$-direction in a brick medium with $T=0.3$~GeV is hadronized by the \texttt{Hybrid Hadronization}. 
As the path length increases, more baryons are produced due to more possible interactions with the medium partons. 
Furthermore, the medium flow shifts the ratio’s peak since baryons inherit more momenta of medium partons. 
%%%%%%%%%%%%%%%%%%%%%%%%%%%%%%%%%%%%%%%%%%%%
%%%%%%%%%%%%%%%%%%%%%%%%%%%%%%%%%%%%%%%%%%%%

\bibliography{main}

\end{document}